\documentclass[prl,aps,reprint,noeprint,superscriptaddress,amsmath, amssymb]{revtex4-1}
\usepackage{graphicx}
\usepackage{subfigure}
\usepackage{lineno}
\usepackage{epsfig}
\usepackage{multirow}
\usepackage{overpic}
\usepackage{color}
\usepackage{bm} 
\usepackage{xspace} 
\usepackage{dcolumn}
\usepackage{booktabs,tabularx}
\usepackage{array}
\usepackage{textcomp}
\usepackage[colorlinks,linkcolor=blue,urlcolor=blue,citecolor=blue]{hyperref}
\usepackage{tikz}
\usepackage{siunitx}
\usepackage{mathrsfs}
\usepackage{amsmath}
\usepackage{cleveref}
\usepackage{float} 

\begin{document}
\normalsize
\parskip=5pt plus 1pt minus 1pt
 \newcommand{\BESIIIorcid}[1]{\href{https://orcid.org/#1}{\hspace*{0.1em}\raisebox{-0.45ex}{\includegraphics[width=1em]{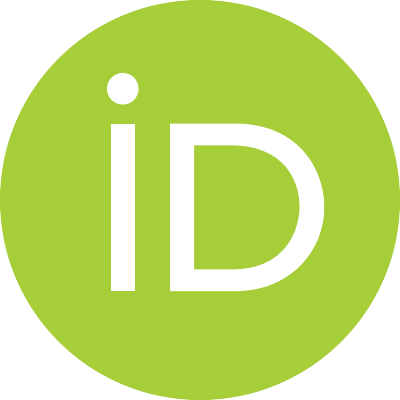}}}} 
\title{
\boldmath 
Observation of $\psi(3770)\to p\bar p$ and Measurement of Electromagnetic Form Factors of Proton at $\sqrt{s} = 3.510-4.946$ GeV
}
\author{BESIII Collaboration}
\thanks{Full author list given at the end of the Letter.}
\begin{abstract}
We search for possible charmonium(-like) states decaying into the $p\bar{p}$ final state by studying the Born cross sections of the $e^+e^-\to p\bar{p}$ reaction, and we determine the proton electromagnetic form factors by analyzing the proton angular distribution. The measurement is performed using a sample of $e^+e^-$ collision data collected at 47 center-of-mass energies from 3.510 to 4.946 GeV, corresponding to an integrated luminosity of 26 fb$^{-1}$, recorded by the BESIII detector collected at the BEPCII collider. The decay $\psi(3770)\to p\bar{p}$ is observed with a significance of 6.6$\sigma$ including systematic uncertainties. 
Furthermore, a structure near 4.2~GeV is observed with significances of $4.6\sigma$ or $4.8\sigma$ for the $\psi(4160)$ or $Y(4230)$ hypotheses including systematic uncertainties, respectively; these interpretations cannot presently be distinguished.
In addition, the moduli of the form factor ratios $|G_{E}/G_{M}|$ and of the magnetic form factors $|G_{M}|$ are extracted by analyzing the proton polar angle distribution with higher precision at large time-like squared momentum transfer. 
These results provide important experimental insights into both the decay mechanisms of charmonium(-like) states in the open charm region and the internal structure of proton.
\end{abstract}

\maketitle
The first member of the charmonium family with $J^{\rm PC}=1^{--}$, the $J/\psi$, was discovered at BNL~\cite{E598:1974sol} and SLAC~\cite{SLAC-SP-017:1974ind} in 1974. Subsequently, a series of charmonium states, such as $\psi(3686)$, $\psi(3770)$, $\psi(4040)$, $\psi(4160)$, and $\psi(4415)$, have been observed\cite{ParticleDataGroup:2024cfk}. Among these, three conventional charmonium states, namely $\psi(4040)$, $\psi(4160)$, and $\psi(4415)$, were identified in measurements of the inclusive hadronic cross section and are dominated by open-charm final states. Their properties are in good agreement with predictions of potential quark models~\cite{Eichten:1974af,Eichten:1978tg,Eichten:1979ms,Barnes:2005pb,Radford:2007vd,Sultan:2014oua}.
In recent years, however, a series of charmonium(-like) states
i.e., $Y(4230)$, $Y(4360)$, $Y(4500)$, $Y(4660)$, and $Y(4710)$,
have been observed through initial-state radiation (ISR) and direct production processes~\cite{ParticleDataGroup:2024cfk}.
These states cannot be easily accommodated within the conventional quark-model framework. As one of the first vector charmonium-like states, $Y(4230)$ remains particularly enigmatic, with several theoretical interpretations proposed to explain its nature~\cite{
Chen:2016qju, Olsen:2017bmm, Guo:2017jvc, Chen:2022asf, Wang:2025clb, Wang:2025dur,Bai:2026atm}. Despite extensive experimental studies, all established decay modes of $Y(4230)$ involve charm hadrons, and no charmless decay channels have yet been confirmed~\cite{ParticleDataGroup:2024cfk}. Investigations of charmless decay modes of $Y$ states therefore provide an important opportunity to discriminate among competing theoretical descriptions and to probe the dynamics of strong interactions in this energy region. 

To date, no observation of baryon–antibaryon decays has been reported for $Y(4230)$ or other charmonium-like states above the open-charm threshold, with the sole exception of the Belle observation of $Y(4660)\to \Lambda_c^+\bar{\Lambda}_c^-$~\cite{Belle:2008xmh}, which remains in tension with subsequent BESIII measurements~\cite{BESIII:2017kqg,BESIII:2023rwv}. To clarify the interplay between baryon-pair production and charmonium(-like) resonances, more precise measurements of the cross sections for exclusive baryonic processes in $e^+e^-$ annihilation are highly desirable. In this context, studies of charmonium(like) decays in the $p\bar{p}$ final state are particularly promising because of the simple event topology and the well-understood angular distribution of outgoing baryons.

Meanwhile, the investigation of the quark–gluon structure of the proton, one of the fundamental building blocks of matter, is of central importance in both nuclear and particle physics. The rich internal structure of the proton originates from Quantum Chromodynamics (QCD), yet remains difficult to calculate in the non-perturbative regime characterized by quark confinement~\cite{Wilson:1974sk} and gluon self-interaction~\cite{Fritzsch:1972jv}. In contrast, the electromagnetic form factors (EMFFs) of the proton can be accessed experimentally in a relatively direct manner~\cite{Hofstadter:1955ae}. These form factors, commonly expressed in terms of the Sachs electric and magnetic form factors $G_E(q^2)$ and $G_M(q^2)$, are Lorentz-invariant functions of the squared four-momentum transfer $q^2$.

Depending on the sign of $q^2$, one distinguishes between the spacelike region, where $q^2<0$ and the form factors can be determined from elastic electron–nucleon scattering, and the timelike region, where $q^2>4M_N^2$ and the moduli of the form factors can be extracted from annihilation processes such as $e^+e^-\to p\bar{p}$. While extensive experimental efforts have been devoted to measurements in the spacelike region, the timelike region remains comparatively less explored~\cite{PACETTI20151}. At large timelike momentum transfers, the small cross sections lead to limited statistical precision, making it challenging to test the asymptotic behavior predicted by perturbative QCD and dimensional counting rules~\cite{PACETTI20151,Brodsky:1973kr}. Previous measurements, for example by CLEO~\cite{Seth:2012nn}, suggest a scaling behavior that deviates from the naive $q^{-4}$ expectation, indicating that the onset of the perturbative regime has not yet been reached. More precise measurements of proton EMFFs at higher timelike momentum transfers are therefore crucial for improving our understanding of nucleon structure and QCD dynamics.

In this Letter, we present a study of decays of charmonium(-like) states into the $p\bar{p}$ final state based on measurements of the Born cross sections for the $e^+e^-\to p\bar{p}$ reaction. The analysis uses $e^+e^-$ collision data corresponding to a total integrated luminosity of 26 fb$^{-1}$, collected at center-of-mass (c.m.) energies ($\sqrt{s}$) between 3.510 and 4.946 GeV~\cite{BESIII:2015qfd,BESIII:2022ulv} with the BESIII detector~\cite{BESIII:2009fln} operating at the BEPCII collider~\cite{Yu:2016cof}. We report the first observation of the decay $\psi(3770)\to p\bar{p}$ and present first evidence for the decays $\psi(4160)\to p\bar{p}$ and $Y(4230)\to p\bar{p}$. In addition, the moduli of the EMFF ratio and of the magnetic form factor $|G_M|$ are determined by analyzing the proton polar-angle distribution with improved precision and at significantly larger timelike squared momentum transfers than previously achieved. More details about the BESIII detector and this analysis can be found in a companion paper submitted simultaneously~\cite{SimultaneousPRDSubmission}.

Candidate events for $e^{+}e^{-}\to p\bar{p}$ are selected using the criteria described in Ref.~\cite{SimultaneousPRDSubmission}. The final state consists of one proton and one antiproton, which are produced back-to-back in the $e^{+}e^{-}$ c.m. system. Positively charged tracks are required to satisfy $E/p<0.5$, where $E$ denotes the deposited energy and $p$ represents the track momentum. To suppress backgrounds, the opening angle between the positive and negative charged tracks is required to satisfy $\theta_{\rm opening}>3.1$ rad, and the hit depth in the muon counter is required to be less than 40 cm. The remaining backgrounds from ISR production of the lower-lying $\psi(3686)$ resonance, Bhabha scattering, and other sources are found to be negligible and to exhibit a smooth distribution in the signal region.

\begin{figure}[!htbp]
    \centering
    \includegraphics[width=0.9\linewidth]{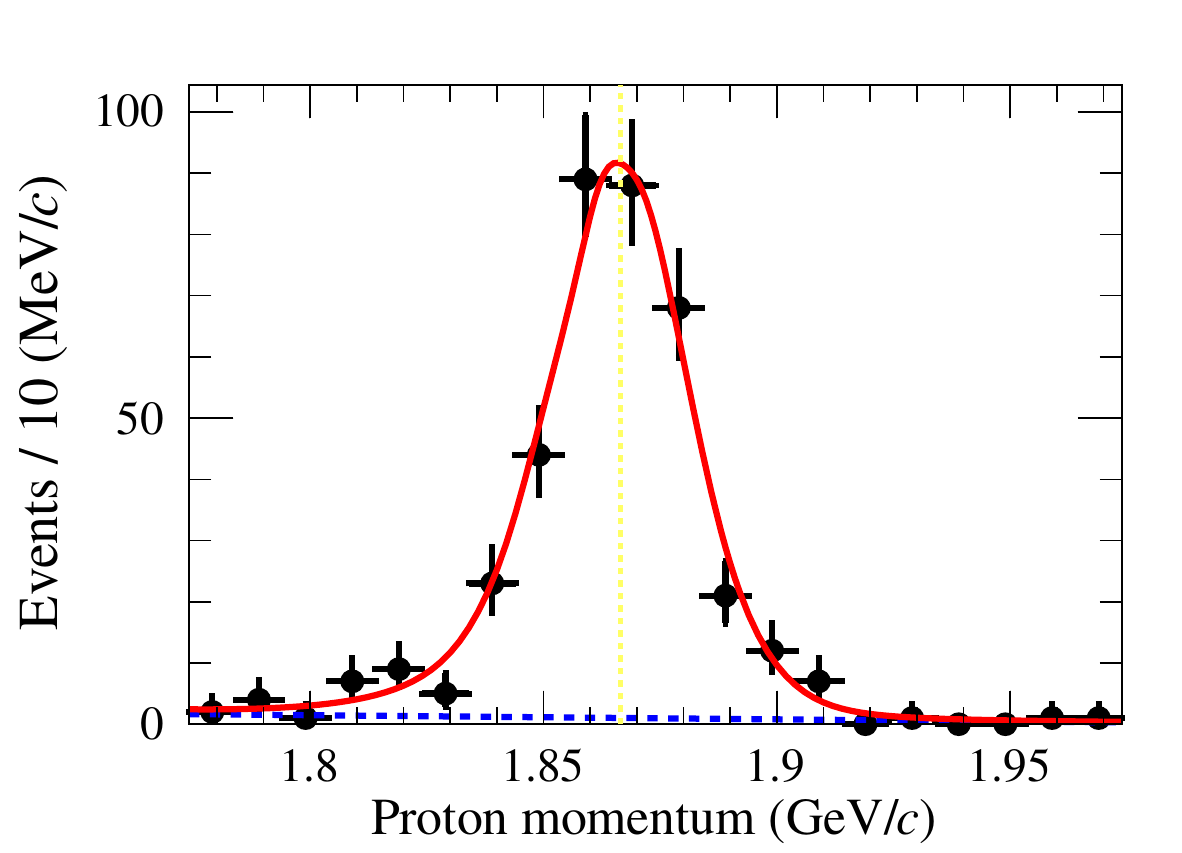}
    \caption{Fit to the proton momentum distribution at $\sqrt{s}=4.178$ GeV. The dots with error bars are data, the red solid line represents the fit result, the dashed line is the background, and the yellow line shows the expected momentum of the proton.}
    \label{4180}
\end{figure}

\begin{figure}
    \centering
    \includegraphics[width=1\linewidth]{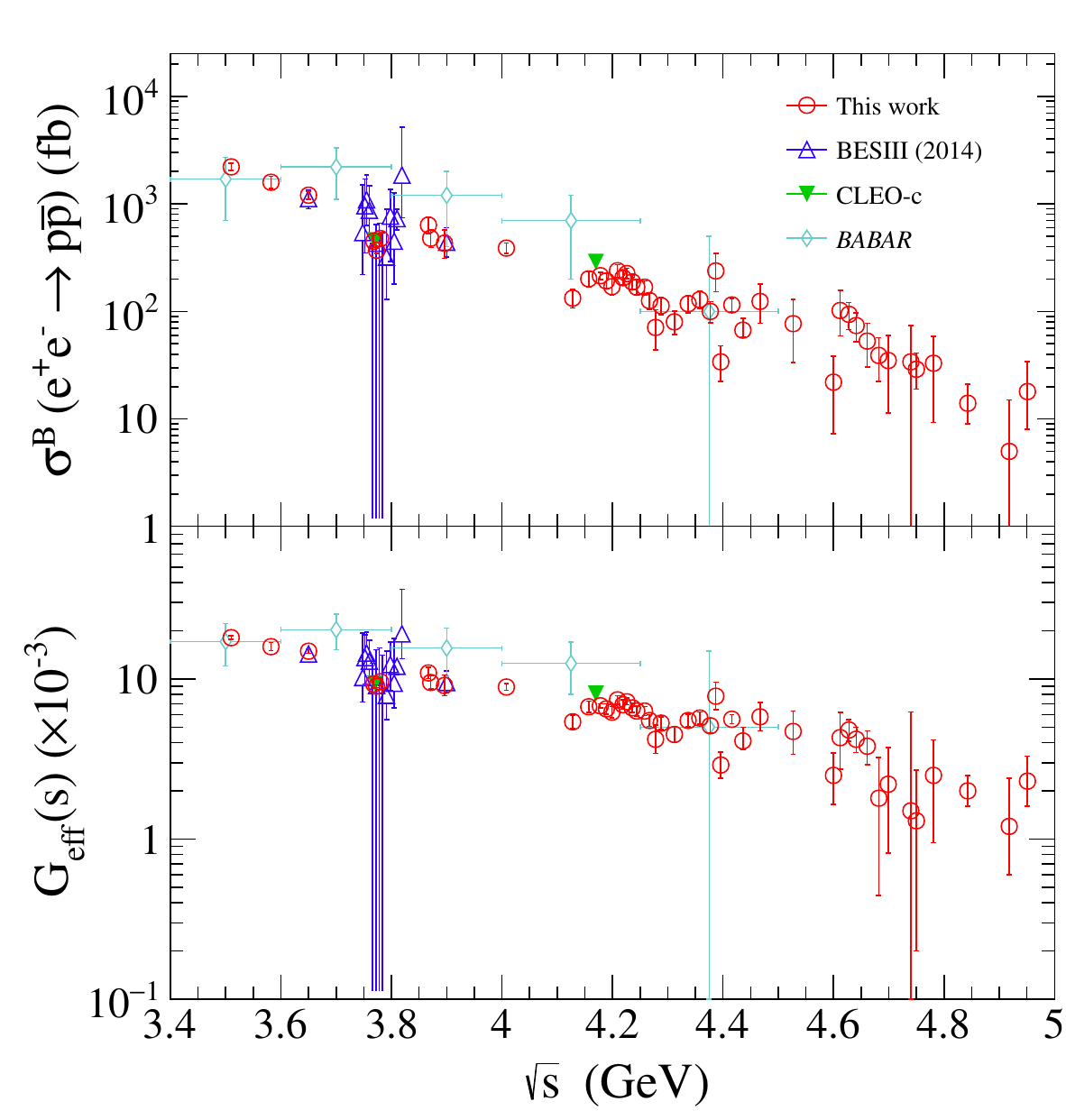}
    \caption{The measured Born cross sections (top) and proton effective
form factors (bottom) for $e^+e^-\to p \bar p$ as a function of the c.m. energy, in comparison to the previous measurements from BESIII~\cite{BESIII:2014fwz}, CLEO-c~\cite{Dobbs:2017hyd}, and $\it BABAR$~\cite{BaBar:2005pon}. The uncertainties include both statistical and systematic contributions.}
    \label{BCS_EFF}
\end{figure}

An extended maximum likelihood fit is used to extract the signal yields for each energy point by fitting the proton momentum distribution as shown in Fig.~\ref{4180}. In the fit, the signal shape for the decay $e^+e^-\to p \bar p$ is represented by the simulated MC shape convolved with a Gaussian function. The background shape is represented by a linear function. More details for the fit to other energy points can be found in Ref.~\cite{SimultaneousPRDSubmission}. 
The efficiencies and ISR correction factors are obtained through an iterative procedure based on the simulated MC sample to accurately measure the Born cross section as proposed in Ref.~\cite{Sun:2020ehv}.
The Born cross sections and effective form factors for the $e^+e^-\to p\bar p$ reaction
in c.m. energies between 3.510 and 4.946 GeV are determined as described in Ref.~\cite{SimultaneousPRDSubmission}. Figure~\ref{BCS_EFF} shows the measured Born cross section and the effective form factor together with previous results.
Our measurements are consistent with earlier results from BESIII, CLEO-c, and BABAR, and achieve improved precision. A clear dip is observed in the mass region of the $\psi(3770)$ resonance.

Potential resonances or charmonium(-like) states produced in the $e^+e^-\to p\bar{p}$ reaction are investigated by fitting the line shape of dressed cross section, $\sigma^{\rm dr}=\sigma^{B}/|1-\Pi|^2$, which includes vacuum-polarization effects through the function $\Pi(s)$. The fit is performed using a least-squares method~\cite{unbais.chisq.epu,unbais.chisq, unbias20031, unbias20032} as defined by
\begin{equation}
\begin{split}
    \chi^2&
    = \sum^{n}_{i}\left(\frac{x_i-fk_i}{\sigma_i}\right)^{2} + \left(\frac{1-f}{\sigma_f}\right)^2.
    \label{chi2}
\end{split}
\end{equation}
Here, $f$ is a normalization factor used to account for correlated contributions, and $\sigma_{f}$ is the relative systematic uncertainty associated with $f$. The quantity $x_{i}$ denotes the measured cross section, $k_{i}$ the expected cross section, and $\sigma_{i}$ the uncorrelated uncertainty at the $i$th c.m. energy. The systematic uncertainties related to the luminosity measurement and to proton tracking and particle identification are assumed to be fully correlated among different c.m. energies, while the remaining systematic uncertainties are treated as uncorrelated. Details of the evaluation of the systematic uncertainties in the cross-section measurements for the $e^{+}e^{-}\to p\bar{p}$ reaction can be found in Ref.~\cite{SimultaneousPRDSubmission}.

 \begin{figure*}[tbp!]
    \centering
    \includegraphics[width=.32\linewidth]{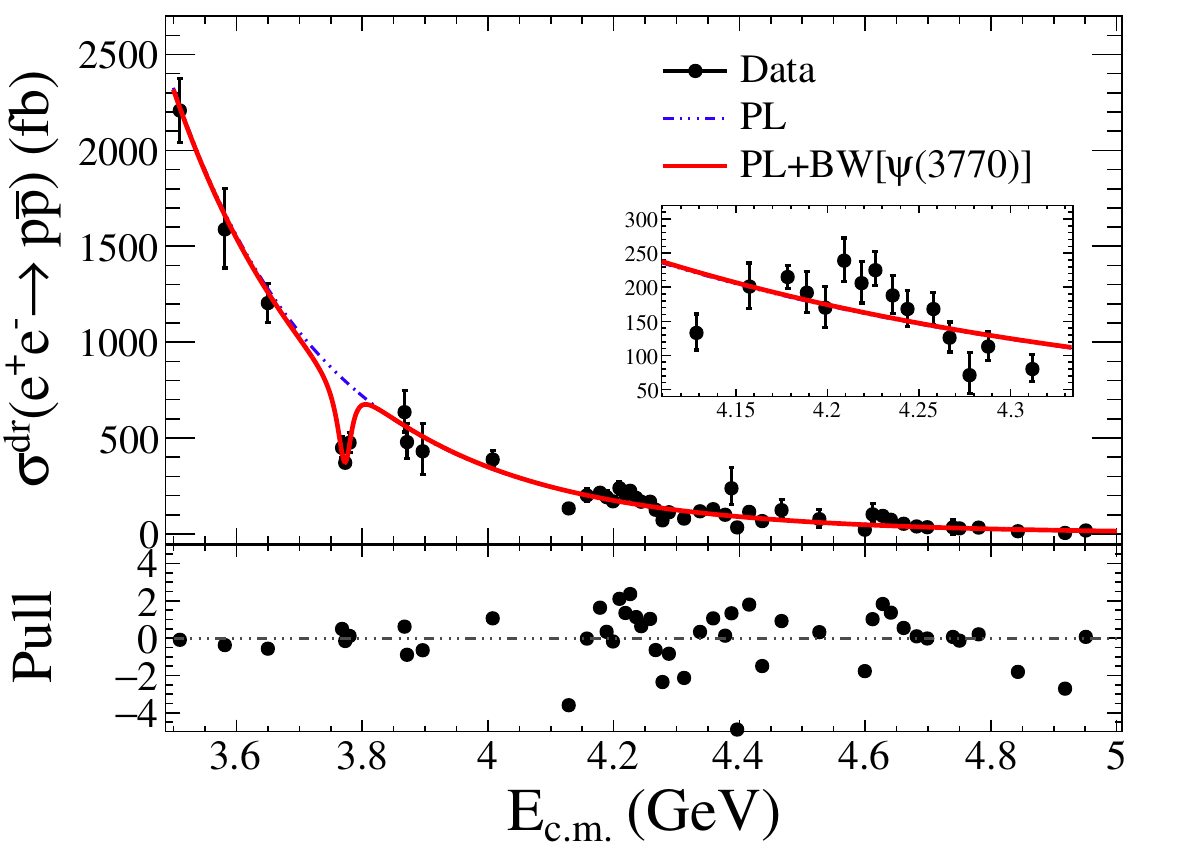}
    \includegraphics[width=.32\linewidth]{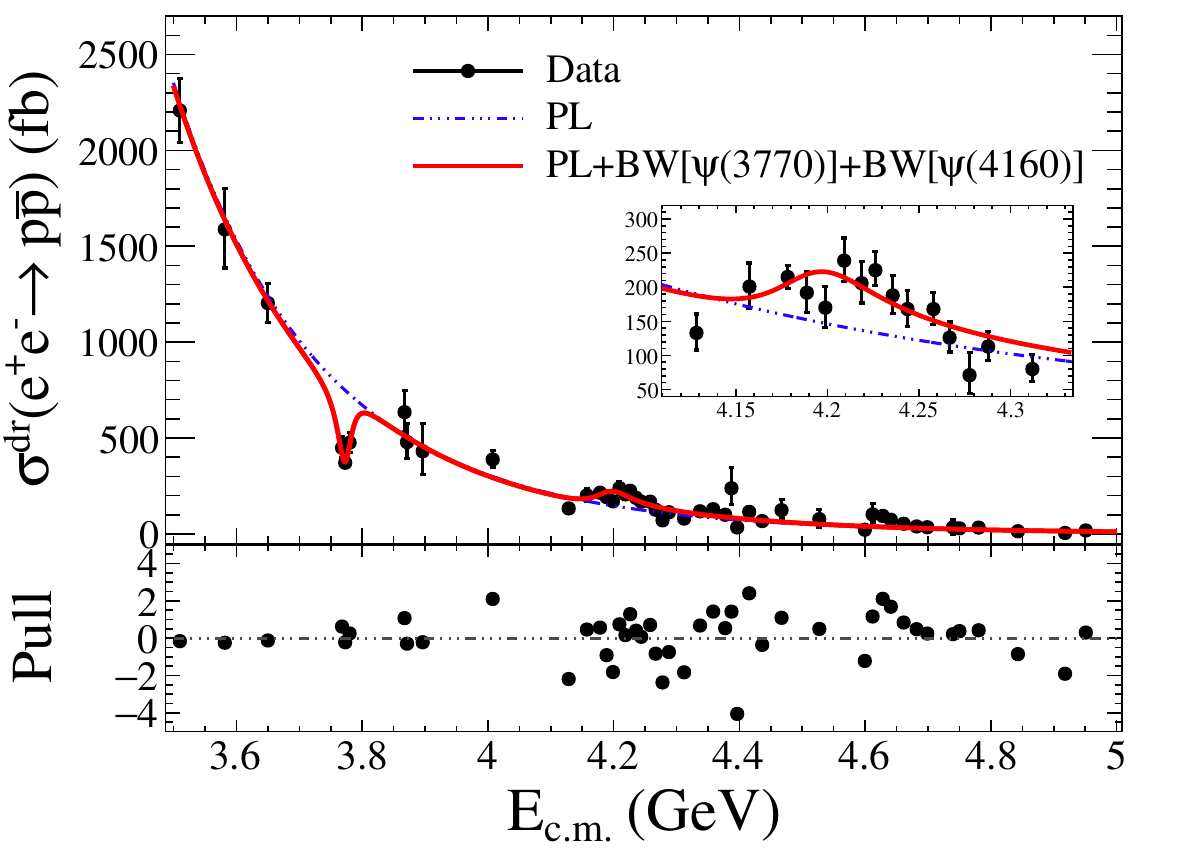}
    \includegraphics[width=.32\linewidth]{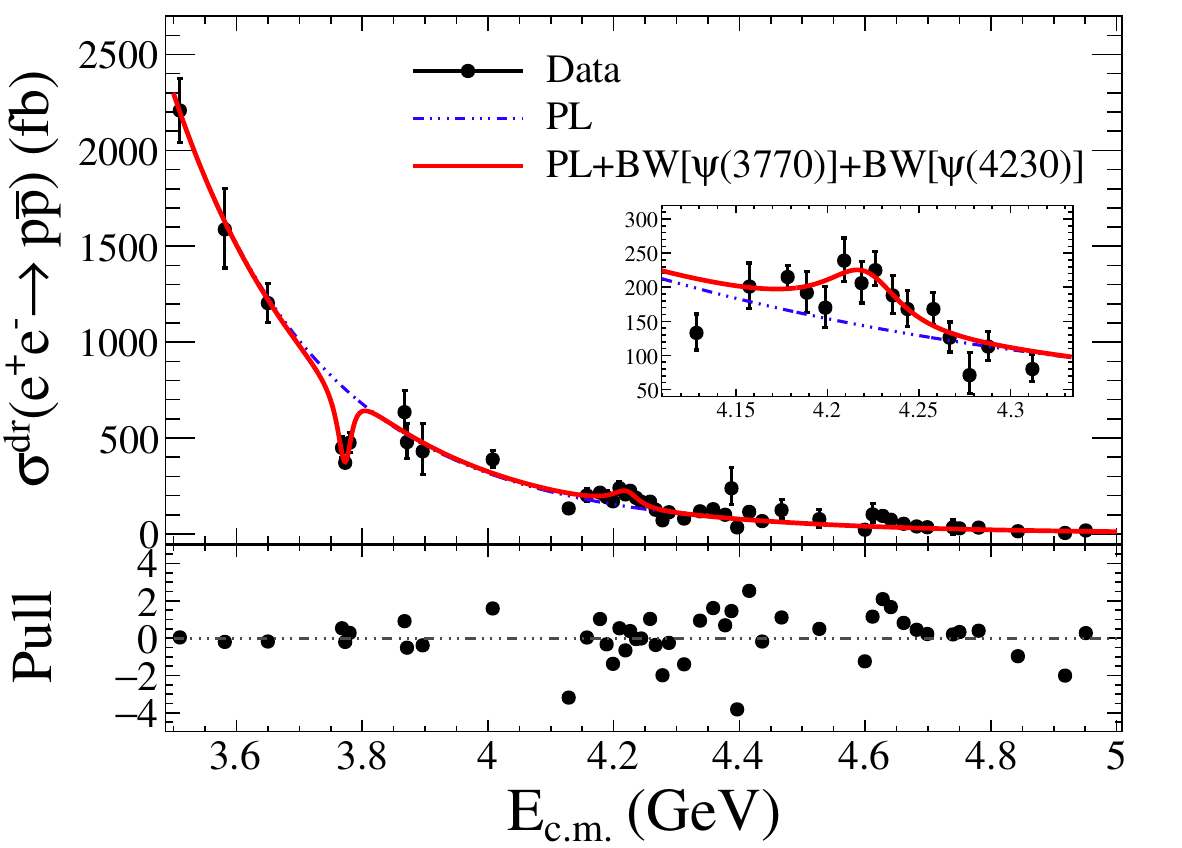}
    \caption{
    Fits to the dressed cross sections for the $e^{+}e^{-}\to p\bar{p}$ reaction at c.m. energies from 3.510 to 4.946 GeV, assuming the presence of the $\psi(3770)$ resonance only, and including an additional charmonium(-like) resonance, either $\psi(4160)$ or $Y(4230)$. The data points with error bars represent the measured dressed cross sections, including both statistical and systematic uncertainties. The blue dot–dashed curves correspond to the fit using only the power-law (PL) function, while the solid red curves show the fit including the resonant contributions. The lower panels display the pull distributions. Insets show enlarged views of the fit in the vicinity of $4.2$ GeV.}
    \label{fit_fit}
\end{figure*}
 \begin{figure*}[tbp!]
    \centering
    \includegraphics[width=.325\linewidth]{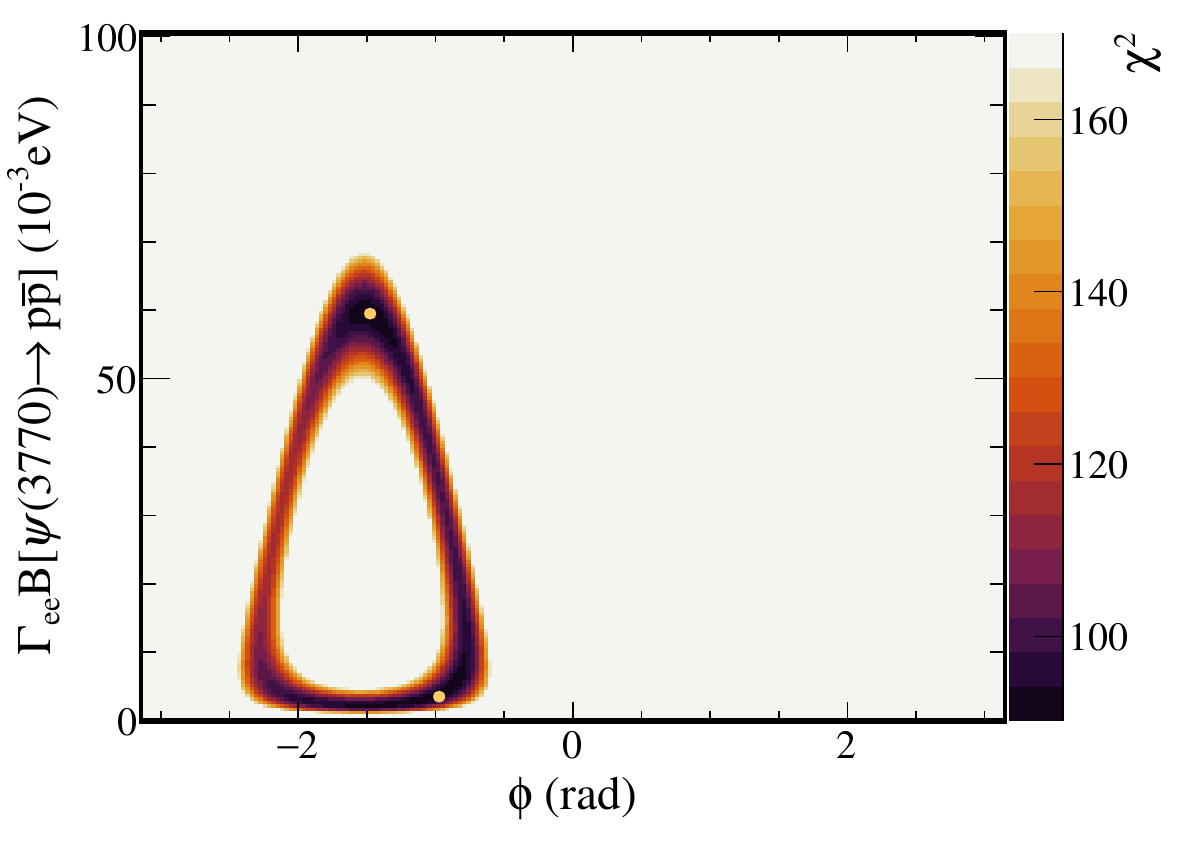}
    \includegraphics[width=.325\linewidth]{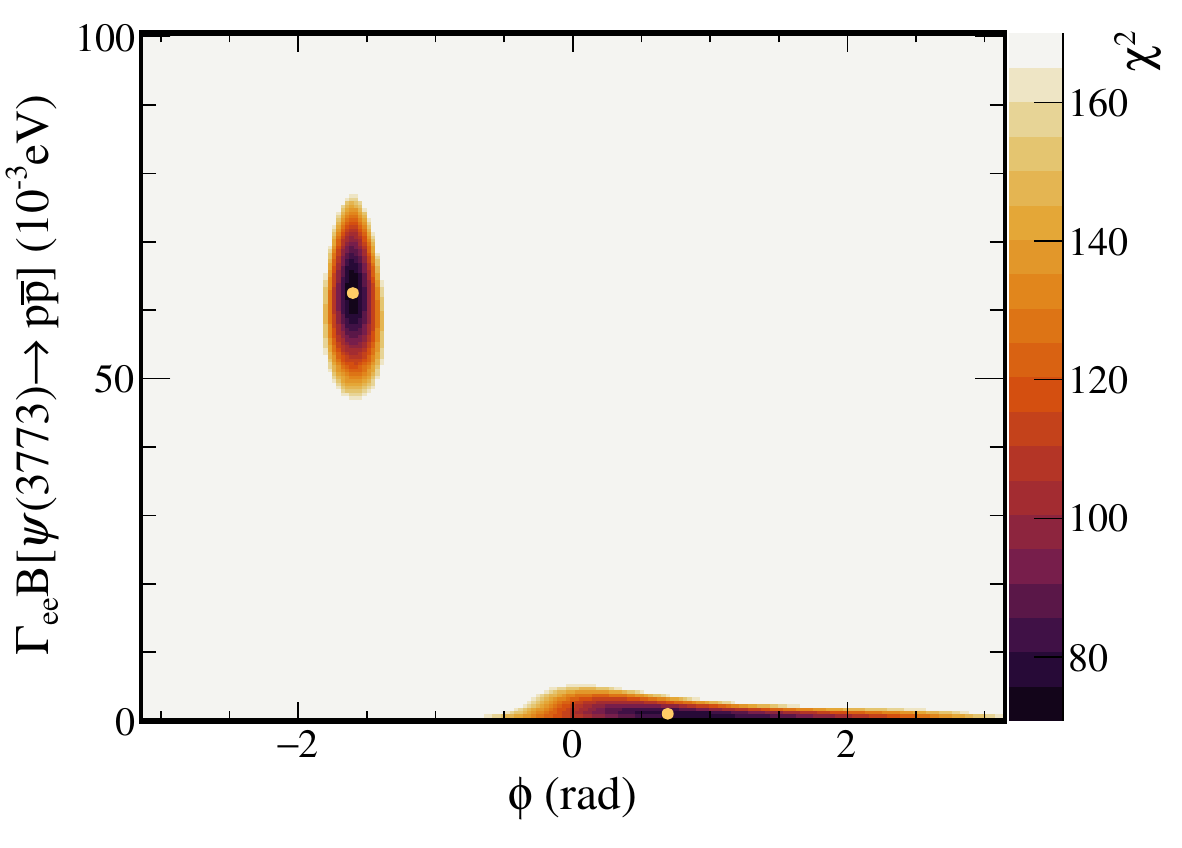}
    \includegraphics[width=.325\linewidth]{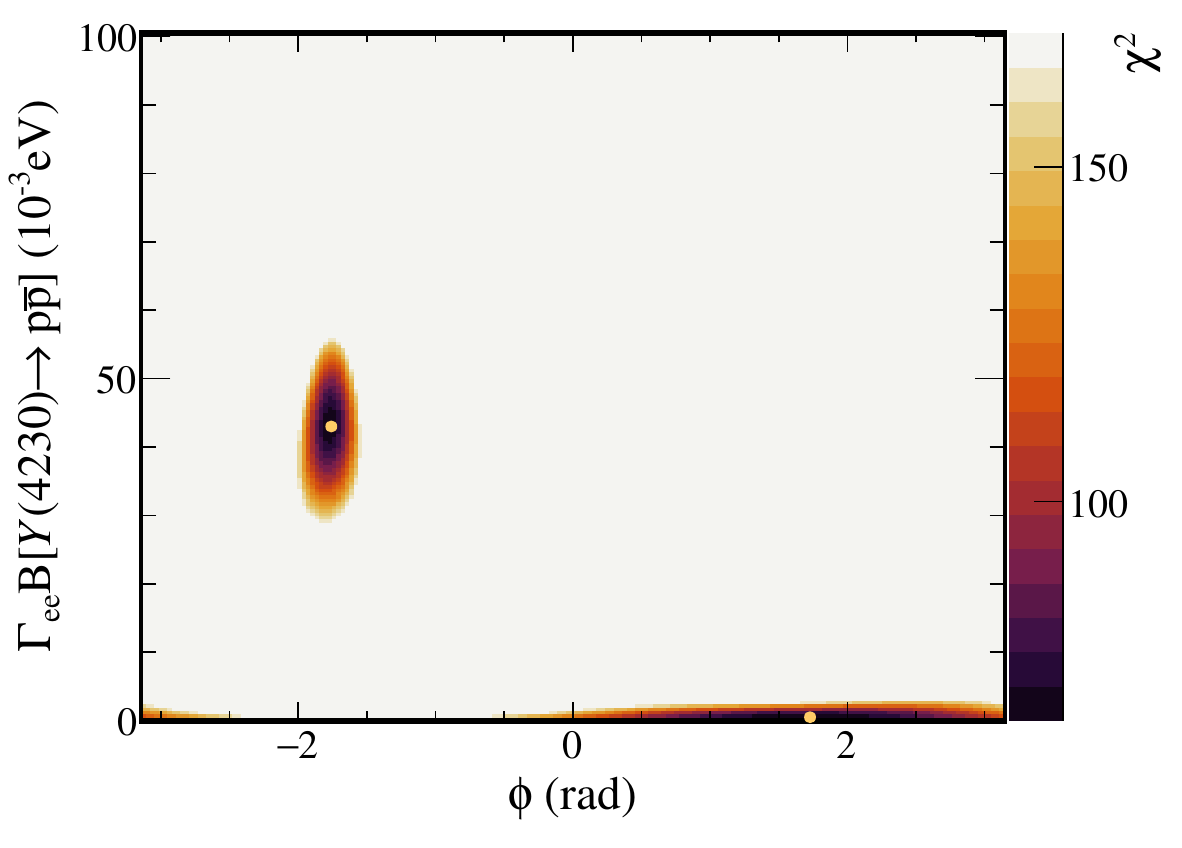}
    \caption{The contour of $\Gamma_{ee}\mathcal{B}$ and $\phi$ on the distribution of $\chi^2$ values for three assumed resonances for $\psi(3770)$, $\psi(4160)$ and $Y(4230)$. The yellow points represent the center values of the best fit.}
    \label{fit_fit:MS}
\end{figure*}
Assuming that the cross section of $e^+e^-\to p \bar p$ contains contributions from a nonresonant continuum, the $\psi(3770)$, and at most one additional resonance, taken in turn from  $\psi(4040)$, $\psi(4160)$, $Y(4230)$, $Y(4360)$, $\psi(4415)$, and $Y(4660)$, the dressed cross section is parametrized as the coherent sum of a power-law (PL) term and two Breit-Wigner (BW) amplitudes,
\begin{equation}\label{BCS}
\begin{split}
    \sigma^{\rm dr}(\sqrt{s}) = \Bigg| \left(\frac{c}{\sqrt{s}}\right)^{\lambda}
    &+ e^{i\phi_1}\rm BW_{\psi(3770)}(\sqrt{s}) \\
    &+ e^{i\phi_2}\rm BW(\sqrt{s}) \Bigg|^{2},
\end{split}
\end{equation}
where $c$ and $\lambda$ are free parameters of the PL term, $\phi_{1}$ and $\phi_{2}$ are the phases of the two resonant amplitudes relative to the PL term,
and the BW function is expressed as
\begin{equation}
    {\rm BW}(\sqrt{s}) =\frac{\sqrt{12\pi\Gamma_{ee}{\cal{B}}\Gamma}}{s-M^{2}+iM\Gamma}.
\end{equation}
The masses $M$ and widths $\Gamma$ of $\psi(3770)$ and the additional resonance are fixed to their PDG values~\cite{ParticleDataGroup:2024cfk}, while all remaining parameters are treated as free. Here, $\Gamma_{ee}\cal{B}$ is the product of the electronic partial width and the branching fraction for the assumed resonance decaying into the $p \bar p$ final state. 
The numerical results for the fitted resonance parameters are summarized in Table~\ref{tab:fit_cs}. The first row corresponds to the fit with the PL term plus the $\psi(3770)$, the remaining rows to the fits in which an additional resonance is included.
Figure~\ref{fit_fit} shows the fits to the dressed cross sections, and Fig.~\ref{fit_fit:MS} presents scans of multiple solutions corresponding to the individual contributions under different resonance hypotheses. Two solutions, including the $\psi(3770)$ together with either the $\psi(4160)$ or the $Y(4230)$ resonance, are obtained and listed in Table~\ref{tab:fit_BCS}. The decay $\psi(3770)\to p\bar{p}$ is observed for the first time with a significance of $6.6\sigma$ including systematic uncertainties. In addition, a clear structure is observed near 4.2~GeV with significances of $4.6\sigma$ or $4.8\sigma$ including systematic uncertainties when assuming the world-average parameters for the $\psi(4160)$ or $Y(4230)$ resonance, respectively. At present, the data do not allow a distinction between these two interpretations. The significance for each assumed resonance is evaluated from the change in $\chi^{2}$ and in the number of degrees of freedom between fits performed with and without the corresponding resonance contribution. The remaining resonances are found to have negligible significance. Owing to the interference between the resonant and continuum amplitudes, multiple solutions with identical $\chi^{2}$ values are obtained~\cite{Bai:2019jrb}.

\begin{table*}
\caption{\label{tab:fit_cs}
The numerical results of resonance parameters from the fits to the dressed cross sections for the $e^+e^-\to p\bar p$ reaction with two solutions (I and II). The fit procedure includes both statistical and systematic uncertainties except for c.m. energy calibration.  The relative phase is given by $\phi$, and $\cal{B}$ denotes the branching fraction of the assumed resonance decaying into the $p\bar{p}$ final state. Note that the values in the brackets represent the upper limit at $90\%$ confidence level. The first and second uncertainties are statistical and systematic, respectively.}
\begin{ruledtabular}
\begin{tabular}{clrlrcc}
& \multicolumn{2}{c}{$\rm Solution\ I$} & \multicolumn{2}{c}{$\rm Solution\ II$} & \multirow{2}{*}{$\chi^2/n.d.f$} & \multirow{2}{*}{$S(\sigma)$} \\
Resonance & $\Gamma_{ee}{\cal{B}}$ ($10^{-3} \rm{eV}$) & $\phi$ $\rm(rad)$ & $\Gamma_{ee}{\cal{B}}$ ($10^{-3} \rm{eV}$) & $\phi$ $\rm(rad)$ & & \\ \hline
$\psi(3770)$ & $59.7\pm3.3\pm 3.3$ & $-1.7\pm0.2\pm 1.0$ & $4.5\pm5.9\pm 2.5$ & $-0.8\pm0.3\pm0.3$ & 2.3 & $6.6$ \\
$\psi(4040)$ & $120.7\pm12.1\pm10.8 \,(<126.8)$ & $0.3\pm0.1\pm0.1$ & $1.2\pm0.6\pm0.1$ & $1.0\pm0.4\pm0.1$ & 2.3 & $1.2$ \\
$\psi(4160)$ & $62.0\pm3.5\pm 13.5$ & $-1.6\pm0.2\pm0.2$ &$1.0\pm0.4\pm0.1$ & $0.7\pm0.2\pm0.1$  & 1.9 & $4.6$ \\
$Y(4230)$ & $44.5\pm1.5\pm6.3$ & $-1.6\pm0.2\pm0.4$ & $0.5\pm0.2\pm0.5$ & $1.7\pm0.2\pm0.2$ & 1.7 & $4.8$ \\
$Y(4360)$ & $54.1\pm4.0\pm0.3\,(<58.6)$ & $-1.8\pm0.1\pm0.2$ & $0.4\pm0.3\pm0.2$ & $-1.8\pm0.3\pm0.1$ & 2.1 & $2.7$ \\
$\psi(4415)$ & $26.2\pm2.2\pm0.1\,(<29.0)$ & $-1.6\pm0.1\pm0.1$ & $0.9\pm0.5\pm0.3$ & $-0.9\pm0.3\pm0.1$ & 2.1 & $2.5$ \\
$Y(4660)$ & $21.0\pm2.7\pm0.1\,(<24.6)$ & $-1.6\pm0.1\pm0.1$ & $0.7\pm0.6\pm0.2$ & $0.4\pm0.6\pm0.0$ & 2.3 & $1.6$ \\
\end{tabular}
\end{ruledtabular}
\label{tab:fit_BCS}
\end{table*}

Systematic uncertainties in the determination of the resonance parameters arise from the uncertainty in the c.m. energy calibration and from assumptions made in the resonance modeling. An uncertainty of 0.8~MeV~\cite{BESIII:2020eyu} is assigned to the c.m. energy and propagated to the resonance masses and widths. The systematic uncertainties associated with multiple solutions are evaluated by comparing the resonance parameters obtained for the nominal solution with those corresponding to the solution that exhibits the largest deviation. Given the significant contributions from the $\psi(3770)$, $\psi(4160)$, and $Y(4230)$ resonances, it is essential to account for uncertainties related to assumptions about resonance interference. These uncertainties are estimated by comparing the resonance parameters extracted under the nominal hypothesis with those obtained from alternative fit models that include additional resonances. Other uncertainties arising from the measurement of the Born cross sections are incorporated directly into the line-shape fit.

In addition, the modulus of the EMFF ratio $R_{EMFF}=|G_E|/|G_M|$ and the modulus of the magnetic form factor $|G_M|$ for the $e^{+}e^{-}\to p\bar{p}$ reaction are determined for the first time using an unbinned maximum-likelihood method based on the analysis of the proton angular distribution in combined energy intervals from 47 data points. Details of this procedure can be found in Ref.~\cite{SimultaneousPRDSubmission}.
\begin{figure}[tbp!]
  \centering
   \includegraphics[width=1\linewidth]{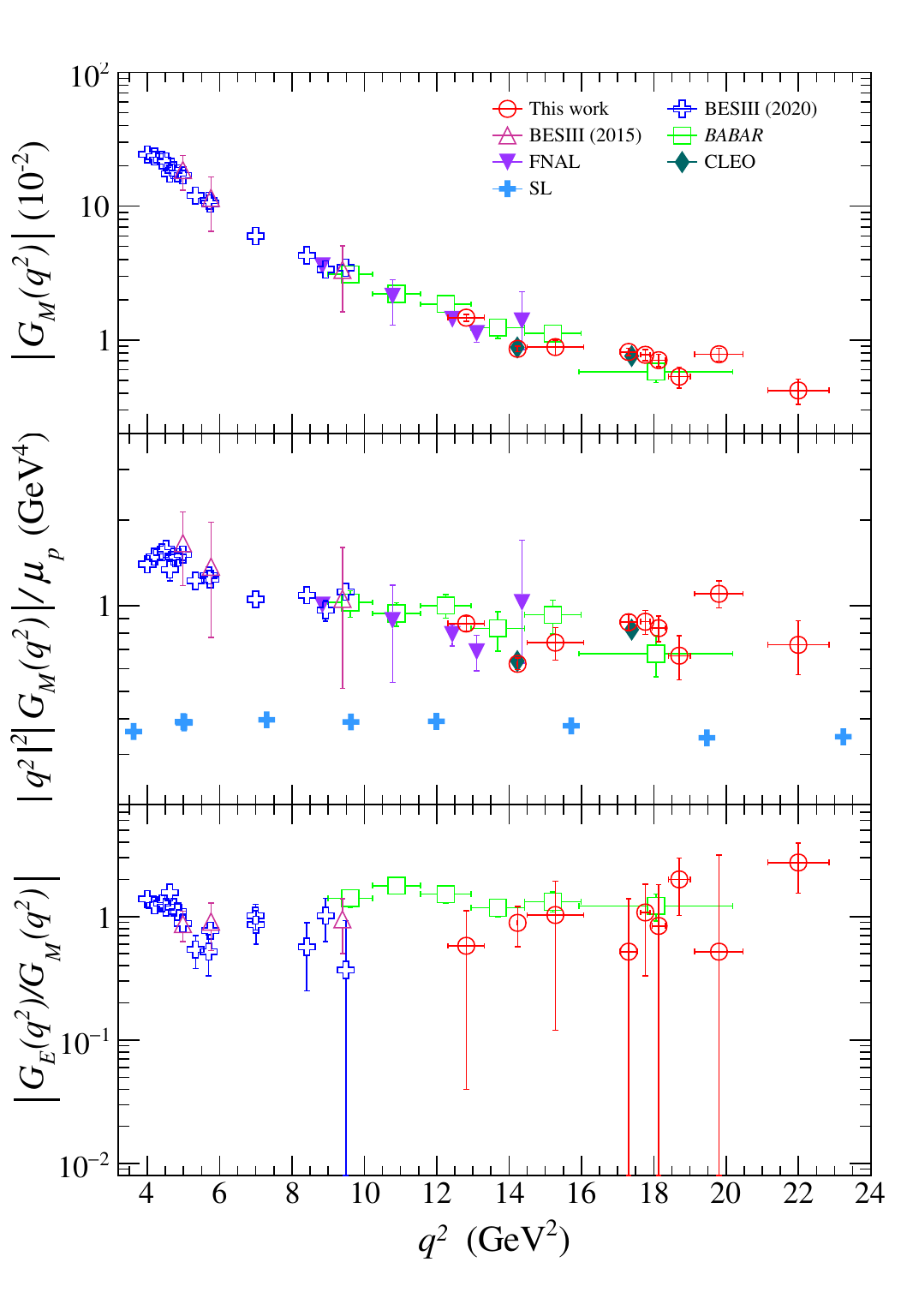}
    \caption{Proton form factors in the $e^+e^-\to p \bar{p}$ reaction as a function of the square of the c.m. energy from this measurement compared to previous measurements at BESIII in 2015~\cite{BESIII:2015axk} (empty triangle) and 2020~\cite{BESIII:2019hdp} (empty cross), {\it BABAR}~\cite{BaBar:2013ukx} (empty square), FNAL~\cite{Andreotti:2003bt} (solid triangle), CLEO~\cite{Seth:2012nn} (solid diamond) and 'SLAC 1993' for the spacelike form factor $G_M$~\cite{Sill:1992qw} (solid square).}
  \label{gm}
\end{figure}
Figure~\ref{gm} shows the measurements of $R_{EMFF}=|G_E|/|G_M|$, $|G_M|$, and $(q^{2})^{2}|G_M(q^{2})|/\mu_{p}$ obtained in this work, compared with previous measurements in the timelike region~\cite{BaBar:2005pon, Andreotti:2003bt, Seth:2012nn,BaBar:2013ukx, BESIII:2015axk, BESIII:2019hdp} for $q^{2}$ values ranging from 3 to 24~GeV$^{2}$. The results are in good agreement with earlier experimental measurements and provide improved precision at large momentum transfers. The determination of $|G_M|$ enables a sensitive test of the expected asymptotic behavior of the proton form factors. For the Sachs EMFFs, dimensional counting rules predict the asymptotic dependence $G_{E,M}(q^{2})\sim(q^{2})^{-2}$~\cite{PACETTI20151}. However, the present results show that $(q^{2})^{2}|G_M(q^{2})|/\mu_{p}=0.64\pm0.04~{\rm GeV}^{4}$ at $|q^{2}|=14.2~{\rm GeV}^{2}$, which is significantly smaller than the values observed at higher momentum transfers. This behavior is consistent with the earlier CLEO measurement~\cite{Seth:2012nn}, which suggested an approximately constant dependence of $q^{2}|G_M(q^{2})|/\mu_{p}$. Given the strong signal observed for $\psi(3770)\to p\bar{p}$, we note that the apparent deviation from the expected asymptotic behavior may be influenced by resonant contributions rather than solely by the running of $\alpha_{s}$. This observation suggests that the running of the strong coupling constant may already play a significant role at these energies. In addition, Fig.~5 includes spacelike data for $|G_M|$ from Ref.~\cite{BaBar:2013ukx}. QCD predicts that the spacelike and timelike form factors should approach the same asymptotic limit as $|q^{2}|\to\infty$~\cite{PACETTI20151}. Our results indicate that the timelike form factors remain significantly larger than the corresponding spacelike values throughout the measured range, extending up to $|q^{2}|=22.0~{\rm GeV}^{2}$.

Systematic uncertainties in the measurements of $R_{EMFF}=|G_E|/|G_M|$ and $|G_M|$ for the $e^{+}e^{-}\to p\bar{p}$ process arise from proton tracking and particle identification (PID), the requirements on the opening angle $\theta_{\rm opening}$ and on the $E/p$ ratio, as well as from background modeling and the fit procedure. The uncertainty related to tracking and PID is estimated by correcting the differences between data and MC simulation and repeating the fit. The uncertainties associated with the requirements on the opening angle and the $E/p$ ratio are evaluated using the Barlow test, which shows no statistically significant deviations; therefore, no corresponding systematic uncertainty is assigned. The uncertainty due to the fit method is estimated through input–output checks based on 500 pseudo-experiments. The differences between the input and output results are taken as systematic uncertainties and are found to be negligible. The total systematic uncertainty is obtained by adding the individual contributions in quadrature. Further details can be found in Ref.~\cite{SimultaneousPRDSubmission}.

In summary, we present an experimental study of charmonium(-like) states decaying into the $p\bar{p}$ final state through precise measurements of the Born cross sections for the $e^{+}e^{-}\to p\bar{p}$ reaction. The decay $\psi(3770)\to p\bar{p}$ is observed for the first time with a significance of $6.6\sigma$ including systematic uncertainties. The corresponding branching fraction is determined to be $(2.27\pm0.23\pm0.22)\times10^{-4}$ or $(0.17\pm0.23\pm0.10)\times10^{-4}$. While the second solution is consistent with expectations based on scaling from the electronic branching fraction, which predicts approximately $0.33\times10^{-4}$~\cite{Dobbs:2017hyd}, the first solution is more than seven times larger. This result indicates that the $\psi(3770)$ resonance must be taken into account when interpreting the CLEO-c data. The observation of $\psi(3770)\to p\bar{p}$ provides the first experimental information on baryonic decays of the $\psi(3770)$.

A clear resonant structure is observed near 4.2~GeV, with significances of $4.6\sigma$ and $4.8\sigma$ for fits assuming a single additional resonance corresponding to either the $\psi(4160)$ or the $Y(4230)$, respectively. However, since the individual contributions of these two resonances cannot be disentangled, the combined significance for the presence of both resonances is $3.6\sigma$. The ratio of branching fractions $R=\mathcal{B}_{Y(4230)\to p\bar{p}}/\mathcal{B}_{Y(4230)\to\pi^{+}\pi^{-}J/\psi}$ is also evaluated using previous BESIII measurements~\cite{BESIII:2016bnd}. Owing to the presence of multiple fit solutions, the largest value, $R=(2.8\pm0.7)\%$, is reported, which is consistent with the upper limit of $13\%$ at the 90\% confidence level obtained by BABAR~\cite{BaBar:2005pon}. This result represents the first experimental evidence for a charmless decay of the $Y(4230)$ and provides important constraints on its underlying structure.

In addition, evidence for the decay $\psi(4160)\to p\bar{p}$ is obtained, corresponding to the first experimental indication of its baryonic decay. The branching fraction is measured to be $(1.31\pm0.66\pm0.71)\times10^{-4}$ or $(0.21\pm0.13\pm0.11)\times10^{-6}$. The ratio $R_{EMFF}=|G_E|/|G_M|$ and the magnetic form factor $|G_M|$ are extracted at several c.m. energies between 3.510 and 4.946~GeV. A plateau-like behavior is observed around $q^{2}\approx14.2~{\rm GeV}^{2}$. Given the strong signal for $\psi(3770)\to p\bar{p}$, this feature may originate from resonant contributions. Furthermore, the timelike form factors remain approximately twice as large as the corresponding spacelike values, in significant contrast with the asymptotic behavior predicted by perturbative QCD at large momentum transfer. These observations indicate that the dynamics governing proton EMFFs are not yet fully understood even at squared momentum transfers as large as $25~{\rm GeV}^{2}$. The present results provide valuable experimental input on baryon-pair decays of charmonium(-like) states and will help constrain theoretical models of baryon structure and production mechanisms.

The BESIII Collaboration thanks the staff of BEPCII (https://cstr.cn/31109.02.BEPC) and the IHEP computing center for their strong support. This work is supported in part by National Key R\&D Program of China under Contracts Nos. 2025YFA1613900, 2023YFA1606000, 2023YFA1606704; National Natural Science Foundation of China (NSFC) under Contracts Nos. 12075107,
11635010, 11935015, 11935016, 11935018, 12025502, 12035009, 12035013, 12061131003, 12192260, 12192261, 12192262, 12192263, 12192264, 12192265, 12221005, 12225509, 12235017, 12342502, 12361141819; 
the Fundamental Research Funds for the Central Universities No. lzujbky-2025-ytA05,  No. lzujbky-2025-it06,  No. lzujbky-2024-jdzx06;
the Natural Science Foundation of Gansu Province No. 22JR5RA389, No.25JRRA799;
the ‘111 Center’ under Grant No. B20063;
the Chinese Academy of Sciences (CAS) Large-Scale Scientific Facility Program; the Strategic Priority Research Program of Chinese Academy of Sciences under Contract No. XDA0480600; CAS under Contract No. YSBR-101; 100 Talents Program of CAS; The Institute of Nuclear and Particle Physics (INPAC) and Shanghai Key Laboratory for Particle Physics and Cosmology; ERC under Contract No. 758462; German Research Foundation DFG under Contract No. FOR5327; Istituto Nazionale di Fisica Nucleare, Italy; Knut and Alice Wallenberg Foundation under Contracts Nos. 2021.0174, 2021.0299, 2023.0315; Ministry of Development of Turkey under Contract No. DPT2006K-120470; National Research Foundation of Korea under Contract No. NRF-2022R1A2C1092335; National Science and Technology fund of Mongolia; Polish National Science Centre under Contract No. 2024/53/B/ST2/00975; STFC (United Kingdom); Swedish Research Council under Contract No. 2019.04595; U. S. Department of Energy under Contract No. DE-FG02-05ER41374.

\bibliography{ref}

\onecolumngrid
\newpage
\begin{center}
M.~Ablikim$^{1}$\BESIIIorcid{0000-0002-3935-619X},
M.~N.~Achasov$^{4,c}$\BESIIIorcid{0000-0002-9400-8622},
P.~Adlarson$^{81}$\BESIIIorcid{0000-0001-6280-3851},
X.~C.~Ai$^{87}$\BESIIIorcid{0000-0003-3856-2415},
C.~S.~Akondi$^{31A,31B}$\BESIIIorcid{0000-0001-6303-5217},
R.~Aliberti$^{39}$\BESIIIorcid{0000-0003-3500-4012},
A.~Amoroso$^{80A,80C}$\BESIIIorcid{0000-0002-3095-8610},
Q.~An$^{77,64,\dagger}$,
Y.~H.~An$^{87}$\BESIIIorcid{0009-0008-3419-0849},
Y.~Bai$^{62}$\BESIIIorcid{0000-0001-6593-5665}
% ...
\\
\vspace{0.2cm}
(BESIII Collaboration)\\
\vspace{0.2cm} {\it
$^{1}$ Institute of High Energy Physics, Beijing 100049, People's Republic of China\\
% ...
$^{4}$ Budker Institute of Nuclear Physics SB RAS (BINP), Novosibirsk 630090, Russia\\
% ...
$^{31}$ INFN Sezione di Ferrara, (A)INFN Sezione di Ferrara, I-44122, Ferrara, Italy; (B)University of Ferrara, I-44122, Ferrara, Italy\\
% ...
$^{39}$ Johannes Gutenberg University of Mainz, Johann-Joachim-Becher-Weg 45, D-55099 Mainz, Germany\\
% ...
$^{62}$ Southeast University, Nanjing 211100, People's Republic of China\\
% ...
$^{64}$ State Key Laboratory of Particle Detection and Electronics, Beijing 100049, Hefei 230026, People's Republic of China\\
% ...
$^{77}$ University of Science and Technology of China, Hefei 230026, People's Republic of China\\
% ...
$^{80}$ University of Turin and INFN, (A)University of Turin, I-10125, Turin, Italy; (B)University of Eastern Piedmont, I-15121, Alessandria, Italy; (C)INFN, I-10125, Turin, Italy\\
$^{81}$ Uppsala University, Box 516, SE-75120 Uppsala, Sweden\\
% ...
$^{87}$ Zhengzhou University, Zhengzhou 450001, People's Republic of China\\

\vspace{0.2cm}
$^{\dagger}$ Deceased\\
% ...
$^{c}$ Also at the Novosibirsk State University, Novosibirsk, 630090, Russia\\
% ...

}
 
\end{center}

\end{document}